\documentclass[11pt,preprint]{aastex}
\begin{document}

\title{Resolving the 47 Tucanae Distance Problem}

\author{Susan M. Percival \& Maurizio Salaris}
\affil{Astrophysics Research Institute, Liverpool John Moores 
       University, Twelve Quays House, Egerton Wharf, Birkenhead CH41 
       1LD, United Kingdom. \newline smp@astro.livjm.ac.uk, ms@astro.livjm.ac.uk}

\author{Francois van Wyk \& David Kilkenny}
\affil{South African Astronomical Observatory, PO Box 9, Observatory
       7935, South Africa \newline fvw@saao.ac.za, dmk@saao.ac.za}

\received{}
\accepted{}

\begin{abstract}

We present new B, V and I-band photometry for a sample of 43 local subdwarfs with {\it HIPPARCOS} 
parallax errors $< 13\%$, in the metallicity range $-1.0 < [Fe/H] < -0.3$, which we use to perform 
main sequence (MS) fitting to the Galactic globular cluster 47 Tuc.  This sample is many times larger 
than those used in previous MS-fitting studies and also enables us to fit in two colour planes, 
$V/(B-V)$ and $V/(V-I)$.
With this enlarged subdwarf sample we investigate whether the current discrepancy in empirical
distance estimates for 47 Tuc, arising from recent MS-fitting and white dwarf fitting results,
is due to inaccuracies in the MS-fitting method.  Comparison of published photometries for 47 Tuc 
has revealed systematic offsets which mean that the $(B-V)$ main-line used in previous studies
may be too blue by $\sim 0.02$ mag, which would have the effect of making any derived distance modulus 
too large by around 0.1 mag.  
Preliminary work has also highlighted discrepancies between results obtained in the 2 colour planes, 
$V/(B-V)$ and $V/(V-I)$.  We have derived main-lines in $V/(B-V)$ and $V/(V-I)$ from the data of 
Kaluzny et al. (1998), which we have recalibrated from the `secondary' standards in 47 Tuc of Stetson 
(2000).  Using an assumed cluster reddening of 
$E(B-V)=0.04$, our best-fit apparent distance modulus is $(m-M)_{V}=13.37^{+0.10}_{-0.11}$ in both 
colour planes, which implies a cluster age of $11.0 \pm1.4$~Gyr and leads to a dereddened distance
modulus of $(m-M)_{0}=13.25^{+0.06}_{-0.07}$.  Comparison with previous work shows that our apparent
distance modulus is $\sim 0.2$ mag smaller than those derived in previous MS-fitting studies.  The 
difference is accounted for by our preferred cluster reddening and the recalibration of the cluster 
photometry, which has made the main-line redder by an average of 0.02 mag in $(B-V)$.  Our derived 
distance modulus is also now plausibly consistent with the short distance recently derived from white 
dwarf fitting.
Independent support for our MS-fitting distance comes from consideration of the Red Clump in 
the cluster, from which we derive a dereddened distance modulus of $(m-M)_{0}=13.31 \pm0.05$, which 
is in agreement with the MS-fitting result.
  
\end{abstract}


\pagebreak 
\section{Introduction}
The age of the globular cluster (GC) 47~Tucanae (NGC 104) plays a fundamental role in the study of 
the formation mechanism of the Galaxy. 47~Tuc belongs to the `thick disk' population of Galactic GCs
and the comparison of its age with that of the more metal poor `halo' GCs and the oldest `thin disk'
open clusters provides vital clues about the timescale for the formation of the various Galactic 
stellar populations (e.g. Salaris \& Weiss 1998, Liu \& Chaboyer 2000, VandenBerg 2000). In 
addition, 47~Tuc provides the zero point for the age determination of `bulge' GCs, since their ages
are most reliably determined from the differential comparison of their Colour-Magnitude Diagrams 
(CMDs) with that of 47~Tuc (Ortolani et al. 1995).  Since the age of a stellar cluster is best 
determined from the comparison of the absolute magnitude of the main sequence turn off stars with 
their theoretical counterpart, the cluster distance modulus must be accurately known.

The two most recent empirical determinations of the 47~Tuc distance provide very different results.
By applying the main sequence fitting technique in the $V/(B-V)$ CMD, using a sample of 7 unevolved
subdwarfs with accurate {\it HIPPARCOS} parallaxes, and adopting $[Fe/H]=-0.7$ and $E(B-V)=0.055$ 
for the cluster, Carretta et al. (2000) have obtained a distance modulus $(m-M)_{V}=13.57 \pm0.09$
(see also Gratton et al. 1997 and Reid 1998, who obtain similar results using small numbers of 
{\it HIPPARCOS} subdwarfs). 
Zoccali et al. (2001) have applied the white dwarf fitting technique in the
$m_{814}/(m_{336} - m_{814})$, $m_{336}/(m_{336} - m_{555})$ and 
$m_{555}/(m_{555} - m_{814})$ CMDs, where  $m_{336}$,  $m_{555}$ and $m_{336}$ correspond roughly 
to the $UVI$ Johnson-Cousins filters; they used a sample of 6 local DA white dwarfs with accurate 
parallaxes, together with 21 WDs identified in 47~Tuc, and determined a distance modulus 
$(m-M)_{V}=13.27 \pm0.14$, adopting $E(B-V)=0.055 \pm0.02$.  The discrepancy between these two 
results is significant, and causes an uncertainty of about 3 Gyr on the cluster age. 

In this work we want to investigate whether this disagreement is due to inaccuracies in the MS-fitting 
distance determination, and provide a new, more solid estimate of 47~Tuc distance 
modulus.   In particular, we aim to greatly improve the reliability of the MS-fitting 
distance by vastly increasing the number of subdwarfs included in the fitting procedure, having an 
homogeneous photometric dataset for all of them, and, for the first time, considering the fitting 
in both the $V/(B-V)$ and $V/(V-I)$ Johnson-Cousins planes.  Because of the different sensitivities
of the $(B-V)$ and $(V-I)$ colours to the stellar metallicities, consistency of the results derived
in these two colour planes is a strong test for the reliability of the distances we obtain. 

In \S 2 we present the photometry of our new large subdwarf sample, and discuss the adopted subdwarf
metallicities, reddenings and Lutz-Kelker corrections; \S3 deals with the 47~Tuc photometry, 
metallicity and reddening, while the MS-fitting results and related systematic errors 
are discussed in \S 4 and \S 5, respectively.  An analysis of the results and comparison with 
other distance determinations appears in \S 6, followed by a summary in \S 7.


\section{The Subdwarf Sample}

\subsection{Selection Criteria}
Suitable subdwarfs were selected using 3 basic criteria - metallicity, parallax error and 
absolute magnitude - as described below.

Metallicities were determined from Str\"{o}mgren indices, available in the literature, via the 
calibration of Schuster \& Nissen (1989).  Likely subdwarf candidates were initially identified from
the catalogues of Schuster \& Nissen (1988), Schuster, Parrao \& Contreras Martinez (1993) and 
Olsen (1993), however, for the sake of consistency, the averaged indices from Hauck \& Mermilliod 
(1998) were used in the analysis.  Metallicities for the subdwarfs were placed on the scale of 
Carretta \& Gratton (1997) (hereafter CG97) using the transformations of Clementini et al. (1999) 
- note that these transformations have a dispersion of $\sim 0.16$ dex in the derived metallicity,
which comes largely from the Schuster \& Nissen calibration.  
Metallicities were restricted to a range within $\pm$0.4 dex of the cluster metallicity 
($[Fe/H]=-0.7$ is assumed for 47 Tuc - see \S 3.2), hence the sample stars all lie in the
range $-1.0 \leq [Fe/H] \leq -0.3$.  This ensures that any uncertainties on the metallicity 
dependence of the main sequence location are kept to a minimum.  2 of the stars in our sample fall just
outside the specified metallicity range when the Hauck \& Mermilliod indices are used (HIP 27080
and HIP 43393 at $[Fe/H]=-0.255$ and -0.252, respectively), therefore these stars were omitted 
from the following analysis.

All stars were required to have {\it HIPPARCOS} parallaxes with errors $\leq 13\%$,
enabling both an accurate determination of their absolute magnitudes and an assessment of 
Lutz-Kelker bias in the sample (see \S 2.3).

The final requirement that $M_{V} > 5.5$ ensures that stars are unevolved, on the lower
main sequence, hence their location in the CMD does not depend on their (unknown) age.

{\it HIPPARCOS} entries for all suitable candidates were carefully checked to avoid the 
inclusion of any suspected binaries or variable stars.  We have identified a sample
of 50 local subdwarfs which meet all the above criteria and we have acquired new B, V and 
I-band photometric data for 43 of them.\footnote{Stars for which we did not manage 
to acquire new photometry are; HIP nos. 1897, 8102, 10798, 56452, 78241, 111209 and 112870} 

\subsection{Observations and Data Reduction}
26 stars in the southern sky sample were observed between January and May 2001 at the Sutherland
site of the South African Astronomical Observatory (SAAO) using the modular photometer on the 
0.5m telescope.  The photometer employs a Hamamatsu R943-02 (GaAs) photomultiplier and a 
Johnson-Cousins $UBV(RI)_{C}$ filter set.  Magnitudes were calibrated using `E-region' standard
stars (see e.g. Menzies et al. 1989) and data reduced using standard SAAO procedures, as described 
fully in Kilkenny et al. (1998) and references therein.  

29 stars in the northern sample were imaged with the CCD on the 1m Jacobus Kapteyn Telescope 
(JKT) on La Palma, on February 9th and June 2nd 2001.  The JKT employs a $SITe 2K$ chip and a
Kitt Peak $BVI$ filter set was used for all observations of programme and standard stars. 
Magnitudes were calibrated from 29 observations of 14 Landolt (1992) standard fields taken over 
the 2 nights.  Basic data reduction was done using standard routines in the {\it FIGARO} data 
reduction package and aperture photometry was performed in {\it GAIA}. 
   
A comparison of the 12 stars in common between the 2 samples shows excellent agreement in the V and I 
filters, the mean differences being 0.006 and -0.002 respectively (SAAO-JKT), with the B-band 
magnitudes displaying a slightly larger mean offset of 0.012 (the JKT magnitudes being brighter).  All 
3 filters have a dispersion of $\sim 0.02$ around the mean, and are therefore consistent with the true 
offsets being zero.  Since several stars in the JKT data have only one observation, the dispersion in 
the mean offset was added in quadrature to the photometry errors for these stars.  
Table 1 presents the new subdwarf data and lists:  {\it HIPPARCOS} number, observed V magnitudes, 
$(B-V)$ and $(V-I)$ colours, photometry errors (in $m$mags), parallax, parallax error, metallicity on 
the CG97 scale, number of observations in each filter and source of photometry (s for SAAO and j for 
JKT).  Figure 1 shows CMDs in $V/(B-V)$ and $V/(V-I)$ for the absolute magnitudes calculated from the 
parallax, and including LK corrections, dividing the subdwarfs into 2 metallicity bins 
($-1.0 \leq [Fe/H] < -0.6$ and $-0.6 < [Fe/H] < -0.25$).  The error bars on the colours are photometry 
errors only. 

\subsection{Lutz-Kelker corrections}
As a consequence of selecting the subdwarfs on parallax error, their absolute magnitudes are 
subject to Lutz-Kelker bias, which leads to a systematic underestimate of their distances (Lutz \&
Kelker 1973).
Lutz-Kelker corrections for the individual subdwarfs were derived from the distribution of proper 
motions for the whole sample, following the procedures of Hanson (1979).  The proper motion 
distribution, identified from the {\it HIPPARCOS} catalogue, was found to be well represented by a 
power law of the form $N(\mu) \propto \mu^{-x}$, where $x = 2.65 \pm0.15$.  The appropriate LK 
correction for each individual star, $\Delta M_{LK}$, is then given by:
$$
\Delta M_{LK} = -2.17 \left[ \left(n + \frac{1}{2}\right) \left(\frac{\sigma_{\pi}}{\pi}\right)^{2} + \left(\frac{6n^{2}+10n+3}{4}\right) \left(\frac{\sigma_{\pi}}{\pi}\right)^{4} 
\right]
$$  
where $n = x+1$ and $(\frac{\sigma_{\pi}}{\pi})$ is the fractional parallax error.

Since the parallax errors for our sample are restricted to $\leq 13\%$, LK corrections 
are small, the mean value being $\Delta M_{LK} = -0.05$ mag.  

\subsection{Reddening}
The Str\"{o}mgren $H\beta$ index offers the best method for assessing the reddening of individual
subdwarfs.  17 stars in our full selected sample have $H\beta$ measurements (Hauck \& Mermilliod
1998) from which $E(b-y)$ was derived using the calibration of Schuster \& Nissen (1989).  
$E(B-V)$ can then be calculated assuming $E(B-V) = 1.35E(b-y)$, and standard extinction laws 
used to determine $E(V-I)$ and $A_{V}$ (e.g. from Cardelli, Clayton \& Mathis 1989). 
We expect reddenings to be very low since the average distance of our sample is only 41pc, the 
furthest lying at 72pc, and in fact, the reddening distribution of these 17 stars peaks at around
$E(b-y)=0.006$ (corresponding to $E(B-V)=0.008$), with tails at higher and lower values (some being 
negative).  
Since the standard error on the Schuster \& Nissen reddening calibration is on the order 0.01 in 
$E(b-y)$, this distribution is consistent with the `true' reddening being zero for 
the sample as a whole.  We note here that these 17 stars include the full distance range of the 
sample, from the nearest to the furthest and the derived reddenings show no correlation with 
distance or Galactic latitude or longitude. 

Several other authors also conclude that reddening effects are negligible in a region within 
75pc of the sun (see, for example, discussions in Perry, Johnston \& Crawford 1982, Blackwell et 
al. 1990 and references therein).  Therefore we chose to adopt zero reddening for the full subdwarf
sample used in our main fits and note that this choice may produce a small systematic effect on the
order of 0.03 mag on the derived distance moduli (see \S 5.1).


\section{47 Tucanae}

\subsection{Photometry}
All previous MS-fitting studies of 47 Tuc have used the $V/(B-V)$ fiducial sequence from 
Hesser et al. (1987).  Since one of the aims of this work is to extend the wavelength range
used for MS-fitting to 47 Tuc we searched for published data which would provide
a well populated lower MS, in both $(B-V)$ and $(V-I)$, extending to $\sim$ 3 magnitudes 
below the turn-off, from which fiducial lines could be derived.  The data of Kaluzny et al.
(1998) seemed ideal for this purpose, as they present B, V and I photometry for more than 22,000
stars covering the whole CMD and extending well down onto the lower MS.  We derived main-lines by
making 0.1 magnitude cuts in V across the main sequence and calculating the mean colour, and the 
dispersion around this mean.  Stars lying more than 1-$\sigma$ from the mean were discarded, and the 
process repeated until the solution converged.  A similar procedure utilizing the mode of the colour 
distribution across the MS, rather than the mean, produced essentially the same results.  Comparison of
the $V/(B-V)$ main-line derived in this way from the Kaluzny et al. data showed it to be in good 
agreement with that of Hesser et al. (see Figure 2). 

Preliminary fits using only the SAAO subdwarf data (26 stars) yielded a $(B-V)$ distance 
modulus in agreement with previous studies (see section \S 6).  However, there appeared to be
a discrepancy of almost 0.3 mag in the derived distance moduli between the $(B-V)$ and 
$(V-I)$ indices (Percival et al. 2001), which prompted us to investigate further 
the consistency of the cluster photometry.  
It should be noted at this point that Hesser et al. (1987) refer specifically
to a possible zero-point offset in their $(B-V)$ data  - `... we believe there is a 
reasonable possibility that our CCD-calibrated scale for 47 Tuc is correct in V but may be 
too blue in $(B-V)$ by $0.01-0.02$ mag'.  Since in the MS fitting procedure any error on the 
colour index is multiplied by the slope on the MS ($\sim 5.5$ in $V/(B-V)$), this 
immediately suggests that any derived distance modulus may be too large by around 0.1 mag.  
Comparison of the Kaluzny et al. data with the photoelectrically calibrated MS data of
Alcaino \& Liller (1987) suggested not only that the $(B-V)$ main-line may be too blue by
$0.02-0.03$ mag, but that the $(V-I)$ main-line may be too red by a similar amount.  
If real, these zero-point offsets combined together would potentially explain the discrepancy
in distance moduli found above.

In order to quantify the suspected zero-point offsets in the Kaluzny data we made a comparison 
of Stetson's `secondary' $BVI_{C}$ standards in 47 Tuc, available through the Canadian 
Astronomy Data Centre web site (http://cadcwww.hia.nrc.ca/standards - see Stetson 2000 for 
details).  A co-ordinate search reliably identified 31 stars in common with the Kaluzny et al. 
data, which spanned the colour range of the main sequence.  Comparison of the photometries 
showed a constant offset in the V filter (and similar constant offset in the $(V-I)$ colour 
index) and a colour dependent offset in $(B-V)$, in exactly the sense suggested above.  
Specifically these derived offsets are:
\newline
$V_{Stetson} = V_{Kaluzny} - 0.025$
\newline
$(V-I)_{Stetson} = (V-I)_{Kaluzny} - 0.026$
\newline 
$(B-V)_{Stetson} = 1.091(B-V)_{Kaluzny} - 0.048$ 
\newline
(see Figure 3).

Applying these offsets, we recalibrated the Kaluzny data from the Stetson standards and rederived 
main-lines in $(B-V)$ and $(V-I)$, as previously described.  The 2 sub-fields for which shorter 
integration times were used (F19 and F10, see Kaluzny et al. 1998 for details) were neglected as these
were clearly not consistent with the remainder of the data.  Fits to our preliminary subdwarf data 
yielded distance moduli that were consistent between the 2 colour indices and so we adopted the
recalibrated main-lines for the remainder of the analysis. 

Figure 2 compares the original Kaluzny main-lines with the recalibrated ones.  The $V/(B-V)$ plot 
also shows the main line of Hesser et al. (1987) used in all previous 47 Tuc MS-fitting studies, 
which is coincident with the original Kaluzny photometry.  Figure 3 illustrates the derived 
photometry offsets between the Kaluzny and Stetson data sets. 

We note here that the $(B-V)$ plot in Figure 3 (middle panel) has one point at $(B-V)=0.57$ which 
appears slightly discrepant and may be forcing the slope of the best-fit line to be too steep.  If
this point is neglected (or given lower weighting) the slope reduces slightly and the best-fit line
becomes: $(B-V)_{Stetson} = 1.071(B-V)_{Kaluzny} - 0.029$.  Adopting this relationship for our
recalibration alters the mean offset by no more than 0.003 mag in $(B-V)$, and the effect on our overall
results is negligible.

\subsection{Metallicity and Reddening}
To enable accurate distance estimates to be made from any method which compares cluster and 
field stars (as is the case using MS-fitting techniques) care must be taken to tie the abundance
scale used for the calibrating local stars to that of the cluster.  Carretta \& Gratton (1997) 
determined metallicities for 24 globular clusters, from cluster giant stars, and find 
$[Fe/H] = -0.70 \pm 0.07$ for 47 Tuc, which we note is in complete agreement with the Zinn 
\& West value of $[Fe/H] = -0.71 \pm 0.08$ and other spectroscopic studies (as discussed in 
Rutledge, Hesser \& Stetson 1997).  Hence, we adopt $[Fe/H]=-0.7 \pm0.1$ for the cluster 
metallicity.  

For the subdwarfs we are using the CG97 scale, as calibrated by Clementini et al. (1999) from the 
Schuster \& Nissen metallicities and, for our fits to be valid, this scale must be consistent with
the metallicity used for 47 Tuc.  
Metallicities for the calibrating subdwarfs used by Clementini et al. are measured according to
procedures which are totally consistent with those used by Carretta \& Gratton (1997) 
to derive the cluster metallicities (see discussion in Clementini et al. 1999), and so the 2 scales
should be comparable.   Reid (1998) also discusses this point and makes a comparison of the CG97
abundances for a sample of 22 local stars which also have spectroscopically derived abundances 
from Axer, Fuhrmann \& Gehren (1994).  Reid finds a mean difference of 0.008 dex in $[Fe/H]$, but 
with most of the offset coming from stars at the lowest metallicities.  However, for 
$[Fe/H] > -1.3$ (which applies to all of our sample, and 47 Tuc) the abundances are found to be in
agreement.

From the above considerations, we make the assumption that the abundance scales we are using
for the subdwarfs and the cluster are self-consistent, and therefore any systematic effects 
should be negligible.  However, we note that any systematic change in the metallicity scales of 
either the subdwarfs or the cluster would affect the derived distances.

Reddening estimates for 47 Tuc from various studies, and using a variety of methods, lie in the 
range $E(B-V)=0.029-0.064$ (see e.g. Crawford \& Snowden 1975, Dutra \& Bica 2000, Gratton et al.
1997 and references therein).  We chose to adopt an $E(B-V)$ value of 0.04 for our main fits, but
explored the range $0.02 \leq E(B-V) \leq 0.06$.  The potential systematic errors associated with
our choice of cluster metallicity and reddening are assessed in \S 5.2.


\section{Main Sequence Fitting}

\subsection{Method}
The basic method employed in this work consists of using the subdwarfs to create a template 
main sequence for comparison with the cluster main-line.  This is done by deriving the 
appropriate metallicity-dependent colour shifts which must be applied to the individual 
subdwarfs in order to `adjust' their metallicity to that of the cluster, $[Fe/H]=-0.7$ 
(noting that decreasing the metallicity of a main sequence star shifts it to bluer colours in
a CMD). 

Employing the $\alpha$-enhanced isochrones of Salaris \& Weiss (1998) for
$[Fe/H]$ = -1.3, -1.0, -0.7, -0.6 and solar-scaled for $[Fe/H]=-0.3$ the procedure is as follows.
For each subdwarf, at its absolute magnitude, interpolate amongst the isochrones to determine 
the {\it theoretical} colour of a star (either $(B-V)$ or $(V-I)$), at ({\it i}) the 
metallicity of the subdwarf and ({\it ii}) the metallicity of the cluster.  The difference 
between these quantities, $\Delta(B-V)$ or $\Delta(V-I)$, represents the colour shift required 
to place the subdwarf at the metallicity of the cluster.  This shift is then applied to the 
{\it observed} colour of the subdwarf, at its absolute magnitude, to build up the template MS. 
Because the metallicity range of our subdwarf sample was restricted to $\pm$ 0.4 dex of the 
cluster metallicity, the derived colour shifts are small.  In $(B-V)$, the mean shift is 0.035 mag
(note that some stars shift to bluer and some to redder colours, depending on the metallicity), and the
maximum shift is 0.050, whilst in $(V-I)$ the mean and maximum shifts are 0.027 and 0.041 respectively.

In order to check whether the colour shifts are dependent on our choice of isochrones,
we carried out tests using different sets of isochrones (e.g. Girardi et al. 2000, the 
diffusive isochrones of Salaris, Groenewegen \& Weiss 2000) and colour transformations (Green, 
Demarque \& King 1987, Castelli 1999).  Since we are utilizing {\it relative} colours for these
shifts, rather than {\it absolute} colours (which can vary significantly from model to model) 
and because we are only adjusting across a narrow metallicity range, the colour shifts produced in 
the different cases were virtually indistinguishable.  Hence we are confident that the 
resulting template MS is not model dependent.

The template MS is then shifted in magnitude, V, and the best fit to the cluster main-line is found 
using a least-squares fitting routine (see \S 4.2 for details).  The required magnitude shift is equal 
to the apparent distance modulus, $(m-M)_{V}$.

Reid (1997) notes that, to be physically correct, the procedure of shifting subdwarfs to the 
cluster metallicity should preserve mass.  This implies that both the colour {\it and} 
magnitude of each subdwarf should be adjusted to be appropriate to a star of equivalent mass, at
the metallicity of the cluster.  The main problem here is that mass also evolves along the MS so
that, at a given colour, luminosity and metallicity, the mass of a star also depends on its age.
Since in general we do not have reliable ages for local subdwarfs it is not possible to apply
these magnitude shifts in a physically correct way.  However, we did test the effect of applying 
colour and magnitude corrections assuming an arbitrary fixed age for the subdwarfs of 10~Gyr.  
The resulting template MS covers a different range of magnitudes than the one which includes colour 
shifts only, as shifting a star of fixed mass to a lower metallicity increases its luminosity, and 
vice-versa.  However, since the shape of the isochrones is very similar across the metallicity range 
we are using, the shape of the resulting template is not changed significantly and we found that fits 
to these revised templates yield the same distance moduli, to within 0.01 mag, as the templates 
constructed using colour shifts only, in both $(B-V)$ and $(V-I)$.

\subsection{Distances from MS-fitting}
Using the whole sample of 41 subdwarfs, `shifted' to $[Fe/H] = -0.7$ (excluding the 2 stars with 
$[Fe/H]>-0.3$), with magnitudes corrected for LK bias and assuming zero subdwarf reddening, our 
best-fit apparent distance modulus is $(m-M)_{V}=13.37 \pm0.03$ from both the $V/(B-V)$ and 
$V/(V-I)$ fits.  The quoted error includes: 
\newline
i) photometry errors for the subdwarfs 
\newline
ii) errors on the subdwarf magnitudes as a result of parallax errors, 
where $\Delta M_{V} = 2.17(\Delta \pi/\pi)$ 
\newline
iii) errors on the subdwarf colours induced by errors on the metallicity calibration, where 
$\Delta[Fe/H]_{subdwarfs}=\pm0.16$ dex (the dominant error here is the dispersion in the Schuster 
\& Nissen calibration from the Str\"{o}mgren indices)
\newline
iv) error on the colour of the cluster main-line resulting from the error on the cluster metallicity 
determination, where $\Delta[Fe/H]_{cluster}=\pm0.1$ dex. 

The best-fit to the cluster main-line is found by using weighted errors for the subdwarfs, and 
minimising $\chi^{2}$.  The fitting routine takes into account errors in both the $x$ and $y$ axes i.e.
errors in both magnitude and colour are accounted for, as detailed above.  
Figure 4 shows the best fits to the recalibrated 47 Tuc main line, with error bars representing the 
errors quoted above.

\section{Testing Our Assumptions}

\subsection{The Subdwarfs;  Metallicity, LK Corrections and Reddening}
In order to test the consistency of the metallicity-dependent colour shifts applied to the 
subdwarfs, we defined a subset of the sample with metallicities in the range 
$-0.9 < [Fe/H] < -0.5$ (13 stars).  The mean metallicity of this subset is $[Fe/H]=-0.64$ and,
since the dispersion on the metallicity calibration is of the order 0.16 dex, we performed fits 
to these stars {\it without} applying any metallicity corrections.  The derived distance moduli 
are in agreement with those derived from the whole (shifted) sample, in both $(B-V)$ and $(V-I)$. 

To check that we are making a correct assessment of the LK bias in the full sample, a subset was 
defined for which $\Delta \pi/\pi < 4 \%$ (implying $| \Delta M_{LK} | < 0.015$) to which LK 
corrections were {\it not} applied (14 stars).  Fits to these stars yield distance moduli in 
complete agreement with the full, LK corrected, sample, hence we are confident that the LK 
corrections are appropriate to the sample as a whole.

As a further consistency check, fits were made to a subset of the sample for which metallicities
are in the range $-0.9 < [Fe/H] < -0.5$ (with a mean of -0.65) {\it and} parallax errors are 
$< 8\%$ (5 stars).  The resultant distance moduli are again in agreement with those from the
whole sample.
As a result of these checks, there do not appear to be any systematic effects arising from 
the metallicity and LK corrections and so we assume an apparent distance modulus of 
$(m-M)_{V}=13.37$ for the remainder of the analysis.

Table 2 lists the best fit distance moduli in $V/(B-V)$ and $V/(V-I)$ for the full sample and
subsets described above, as follows:
\newline
1 - Full sample, LK corrected magnitudes, `shifted' to $[Fe/H] = -0.7$
\newline
2 - Subset with metallicities in the range $-0.9 < [Fe/H] < -0.5$ (mean $[Fe/H]=-0.64$), fitted
{\it without} applying metallicity corrections
\newline
3 - Subset with parallax errors $< 4 \%$, no LK corrections applied 
\newline
4 - Subset with metallicities in the range $-0.9 < [Fe/H] < -0.5$ {\it and} parallax errors 
$< 8 \%$

All the above tests were made assuming zero reddening for the subdwarfs.  However, if we apply 
reddenings derived from the $H\beta$ index to the stars for which this is available and use an 
average reddening of $E(B-V) = 0.008$ for the rest of the sample (see \S 2.4), distance moduli are
reduced by $\sim 0.03$ mag in all the fits. 

The referee expressed concern about the possibility of a metallicity bias in our subdwarf sample which 
may arise due to the underlying metallicity distribution of the parent population.  Since the 
metallicity distribution of local subdwarfs is strongly skewed towards higher metallicities, this would
cause the metallicity of individual subdwarfs to be underestimated and make MS-fitting distances appear
too long.  We took the underlying `true' metallicity distribution to be represented by the combined 
samples of Schuster \& Nissen (1988) and Schuster, Parrao \& Contreras Martinez (1993), since most of 
our sample was drawn from these catalogues.  There is a large peak in this distribution at 
$[Fe/H]_{SN} \sim -0.2$ due to disk stars, and another smaller peak at $[Fe/H]_{SN} \sim -1.3$ due to 
the halo population (see Schuster, Parrao \& Contreras Martinez 1993, their Figure 1).  
To quantify the effect on our sample, we made Monte Carlo simulations of the metallicity 
distribution of the parent population, as described above, combined with random errors, generated 
assuming a Gaussian error distribution of width 0.16 dex (Schuster \& Nissen 1989).  
Then, at the observed metallicity of each individual subdwarf in our sample, we considered all the
stars within 0.16 dex of this value, and calculated the difference between the means of the `observed' 
metallicities (after the Gaussian errors have been added) and the `true' metallicities (from the 
initial distribution).  The difference between these quantities provides an estimate of the likely bias
in our sample.  Offsets vary between 0.021 and 0.043 dex with a mean of 0.035 dex, in the sense that the
`true' metallicities are higher.  Applying individual metallicity corrections to our subdwarfs and 
refitting the sample to the 47 Tuc main-line results in a distance modulus which is smaller by only 
0.02 mag.    

\subsection{The Cluster;  Reddening and Metallicity}
Reddening for the cluster was taken to be $E(B-V)=0.04$ in all our fits.  Increasing or 
decreasing $E(B-V)$ by 0.02 mag changes the apparent distance modulus, $(m-M)_{V}$, by 
$\pm 0.1$ in the sense that larger assumed reddening produces a longer distance.  Since reddening
values for 47 Tuc available in the literature are not necessarily in agreement within their quoted
errors and come from very disparate methods, we have no way of truly assessing the appropriate 
reddening to apply or of calculating the associated errors, in a statistical sense.  
Conservatively, we have chosen to adopt $\pm 0.1$ as the reddening-induced error on the apparent 
distance modulus, although this undoubtedly represents more than a $1\sigma$ error.
Note, however, that this uncertainty of $\pm 0.1$ on the apparent distance modulus, $(m-M)_{V}$, only
induces an uncertainty on the true distance modulus, $(m-M)_{0}$, of $\pm 0.05$.

Our final derived apparent distance modulus for 47 Tuc, including the uncertainties on cluster and 
subdwarf reddening in the total error budget, is:
$(m-M)_{V} = 13.37^{+0.10}_{-0.11}$, leading to a dereddened distance modulus of 
$(m-M)_{0} = 13.25^{+0.06}_{-0.07}$.

As already discussed, the metallicities we are using for both the cluster and the subdwarfs are
consistent with the CG97 scale and therefore we expect there to be no associated
systematic errors.  However, it should be borne in mind that there is an inherent uncertainty in 
any study of this kind, since there is always the implicit assumption that the cluster metallicity,
generally obtained from giant stars, is the same as that which would be obtained from the 
cluster main sequence.  Coupled with the still controversial zero-point for the absolute temperature 
scale of giant stars, it is possible that a systematic shift in the metallicity scale may become 
necessary in the future.  Changing the metallicity scale of either the subdwarfs or the cluster, 
with respect to one another, would modify the distance moduli derived from our subdwarf sample by 
$\sim 0.1$ mag for each 0.1 dex change in metallicity, in the sense that decreasing the 
metallicity of the cluster would {\it decrease} the distance, whilst decreasing the metallicity of
the subdwarfs would {\it increase} it.  We note here that if we change the {\it relative} metallicities
by more than $\sim$ 0.2 dex, the distance moduli we derive from the 2 colour indices are no longer 
consistent within their errors.

Another cause for concern may be the variation in C and N abundances which are known to exist in 
47 Tuc.  However, the study of Cannon et al. (1998), which measures abundance variations down onto
the main sequence of 47 Tuc, indicates that these are unlikely to affect the broad-band colours 
used in MS-fitting studies (see their Figure 6).


\section{Discussion}

Employing our distance determination discussed in the previous section we then determined 
the age of the cluster.  The luminosity of the turn-off was estimated by fitting a 
parabola to stars in the magnitude range $17.3 < V < 18.0$.  This yielded a turn-off magnitude of 
$V_{TO} = 17.66 \pm 0.1$ which, coupled to our distance modulus and assuming $[Fe/H]=-0.7$, 
provides an age of $11.0\pm 1.4$ Gyr when using the Salaris \& Weiss (1998) isochrones
(for $[Fe/H]=-0.7$, $Y=0.254$, $[\alpha/Fe]=0.4$). The error bar on the age takes into account the
error on the cluster distance and the error on the apparent magnitude of the turn off.

For direct comparison with previous MS-fitting work we followed Carretta et al.~(2000) by fitting 
our whole subdwarf sample (LK, reddening and extinction corrected) to the $V/(B-V)$ main-line 
of Hesser et al. (1987), using a cluster reddening of $E(B-V)=0.055$.  The best fit yielded an 
apparent distance modulus of $(m-M)_{V}=13.59 \pm0.03$ (errors as in \S 4.2 only), which is in 
excellent agreement with the Carretta et al.~(2000) result of $13.57 \pm0.09$.  The reason for the 
lower distance modulus we have obtained in the previous section is due to both our preferred cluster 
reddening, and the recalibration of the 47~Tuc photometry.
A reduction of $\sim$0.1 mag comes from the different cluster reddening, while the rest of the 
difference is due to the photometry recalibration, which makes the main-line redder, on average, by
$\sim 0.02$ in $(B-V)$ (see Figure 2 with the comparison of the main-lines). 

To investigate whether our MS-fitting distance is still significantly different from that derived 
from white dwarf fitting, we determined the 47~Tuc distance modulus with our recalibrated cluster 
photometry and a reddening E(B-V)=$0.055\pm 0.02$, as used in Zoccali et al.~(2001). We obtain
$(m-M)_{V}=13.45\pm0.10$ which has to be compared with $(m-M)_{V}=13.27\pm0.14$ obtained from the
white dwarf-fitting method.
The two results are less discrepant than in the case of the Carretta et al.~(2000) main sequence 
distance, especially if one takes into account the possibility of an additional systematic error 
of the order of 0.1 mag on the white dwarf distance, due to the uncertainty on the value of the 
mass and on the thickness of the hydrogen envelopes for the cluster white dwarfs (see the detailed
discussion in Salaris et al.~2001).

Independent support for our main sequence distance determination comes from the use of the Red 
Clump (RC) as a distance indicator.  The red horizontal branch of 47~Tuc is the counterpart of the
RC in the solar neighbourhood discussed by Paczyinski \& Stanek~(1998), whose I-band absolute 
magnitude is very precisely determined by {\it HIPPARCOS} parallaxes as $M_{I,RC}= -0.23\pm$ 0.03 
(Stanek \& Garnavich~1998). Girardi \& Salaris~(2001) have discussed in great depth the use of the
RC as standard candle and provide evolutionary corrections to the absolute brightness of the
local RC, which take into account the effect of the star formation history and age-metallicity 
relationship of the stellar population under scrutiny. In the case of a GC like 47~Tuc, the Girardi
\& Salaris~(2001) method predicts a correction of only +0.03 to the absolute $I$-band brightness 
of the local RC.
From the recalibrated photometry of Kaluzny et al.~(1998) we obtain $I_{RC}$=13.18$\pm$0.02 which,
together with $M_{I,RC}= -0.20 \pm0.03$ appropriate for the 47~Tuc RC, provides a distance modulus
of $(m-M)_{I,RC}=13.38\pm0.04$. Considering a reddening $E(B-V)=0.04\pm 0.02$ and the extinction 
law by Cardelli et al.~(1989), one gets a dereddened distance modulus from this method of 
$(m-M)_{0,RC}=13.31\pm0.05$, which agrees well with that obtained from the main sequence fits, 
$(m-M)_{0}=13.25^{+0.06}_{-0.07}$.

\section{Summary}                                                                          
We have defined a sample of 50 local subdwarfs in the metallicity range $-1.0<[Fe/H]_{CG97}<-0.3$,
which are suitable for main sequence fitting to the GC 47~Tuc.  We have obtained new photometric 
data in B, V and I for 43 subdwarfs in the sample and use these the derive the 47~Tuc distance 
modulus in the $V/(B-V)$ and $V/(V-I)$ planes.  

From a careful comparison with the 47~Tuc standards of Stetson (2000), we have found that the 
commonly used Hesser et al. (1987) main-line in $V/(B-V)$ appears to be too blue by an average of 0.02 
mag.  This implies that any distance modulus derived via MS-fitting from the Hesser main-line would be 
too large by $\sim 0.1$. 
In order to perform MS-fitting in 2 colour indices ($(B-V)$ and $(V-I)$), we have derived main-lines 
from the photometry of Kaluzny et al. (1998), which we have recalibrated using the Stetson (2000) 
standards.  

Theoretical isochrones are used to calculate the metallicity-dependent colour shifts required to 
move the subdwarfs in the CMD to match the metallicity of the cluster, and hence build up a template 
MS.  However, we find that the particular choice of isochrones does not influence the 
magnitude of the derived colour shifts and so the method does not appear to be model dependent. 

We have investigated the effects of metallicity scales, LK bias and reddening, for both
the subdwarfs and the cluster, on the derived distance moduli.  We find no appreciable errors arising
from either the assumption of zero reddening for the subdwarfs or from our assessment of the LK 
bias in the sample.  The largest single source of error is the uncertainty on the cluster reddening 
which propagates through to an error on the order of 0.1 mag on the distance modulus.
Metallicities used for both the
cluster and the subdwarfs are consistent with the CG97 scale and therefore there should be no 
associated systematic errors.  However, any change in zero-point of the metallicity scale which 
altered the cluster or subdwarfs with respect to each other, would potentially alter the distance 
moduli by $\sim 0.1$ mag for every 0.1 dex change in the metallicity.   

Our best-fit apparent distance modulus for 47~Tuc is $(m-M)_{V}=13.37^{+0.10}_{-0.11}$, with an assumed 
cluster reddening of $E(B-V)=0.04$.  Coupled with our assessment of the apparent magnitude of the 
turn-off ($V_{TO}=17.66\pm0.1$), the implied age of the cluster is $11\pm1.4$ Gyr.  
The dereddened distance modulus is $(m-M)_{0}=13.25^{+0.06}_{-0.07}$.   

The apparent distance modulus we find for 47~Tuc is $\sim 0.2$ mag smaller than those derived in
previous MS-fitting studies.  This is due partly to our preferred cluster reddening and partly to
the recalibration of the 47~Tuc photometry, which has made the main-line redder than that used by 
all previous studies, by an average of 0.02 mag in $(B-V)$.
Our MS-fitting distance is now plausibly consistent with that derived from white dwarf fitting.

Independent support for our MS-fitting distance is given by consideration of the Red Clump in the 
cluster, from which we derive a dereddened distance modulus of 
$(m-M)_{0,RC}=13.31\pm0.05$, which is in good agreement with that derived from MS-fitting.  

\acknowledgements{
We thank the anonymous referee for a thorough reading of the manuscript and for useful comments and 
suggestions.
We also thank Anna Piersimoni and Giuseppe Bono for helpful discussions and Phil James for careful 
reading of the manuscript at various stages, which has helped to improve the clarity of this paper.
SMP acknowledges financial support from PPARC. 

SMP and MS are Guest Users of the Canadian Astronomy Data Center, which is operated by the Dominion 
Astrophysical Observatory for the National Research Council of Canada's Herzberg Institute of 
Astrophysics. 

This research has made use of the {\it SIMBAD} database, operated at CDS, Strasbourg, France.}



\begin{deluxetable}{rrccrrrccccc}
\tablecolumns{12}
\tablewidth{0pc}
\tablecaption{New Photometry for Subdwarf Sample}
\tablehead{
\colhead{} HIP &  \multicolumn{1}{c}{V} & $(B-V)$ & $(V-I)$ & $\sigma_{V}$ & $\sigma_{BV}$ & $\sigma_{VI}$ & $\pi$ & $\Delta\pi$ & $[Fe/H]_{CG}$ & {\it N} & S}
\startdata
   5004 & 10.250 & 0.752 & 0.869 &  9 &  0 &  4  & 16.28 & 1.76 & -1.000 & 2 & s  \\
  15126 & 10.232 & 0.694 & 0.774 & 21 &  0 & 16  & 12.64 & 1.66 & -0.707 & 2 & sj \\
  23431 &  8.234 & 0.720 & 0.838 & 23 & 32 & 19  & 34.88 & 1.46 & -0.379 & 2 & j  \\
  27080 &  8.064 & 0.772 & 0.848 &  5 &  3 &  0  & 39.08 & 0.75 & -0.255 & 3 & s  \\
  28898 &  8.427 & 0.854 & 0.940 &  6 &  3 &  2  & 36.53 & 0.74 & -0.342 & 3 & s  \\
  33940 & 10.150 & 0.827 & 0.982 & 25 & 34 & 22  & 17.19 & 2.28 & -0.347 & 3 & j  \\
  33982 &  9.470 & 0.631 & 0.789 & 25 & 34 & 22  & 16.48 & 1.76 & -0.549 & 3 & j  \\
  43393 &  9.166 & 0.742 & 0.808 & 13 &  6 &  4  & 18.78 & 1.48 & -0.252 & 3 & sj \\
  44176 &  9.691 & 0.900 & 1.093 & 24 & 33 & 20  & 25.33 & 1.60 & -0.378 & 3 & j  \\
  47033 &  8.867 & 0.715 & 0.765 & 23 & 34 & 19  & 21.43 & 1.39 & -0.340 & 3 & j  \\
  47961 &  9.435 & 0.711 & 0.756 &  4 &  9 &  4  & 16.28 & 1.57 & -0.323 & 3 & sj \\
  51127 &  9.750 & 0.977 & 1.119 &  5 &  4 &  1  & 32.50 & 1.62 & -0.419 & 3 & sj \\
  51769 & 10.486 & 0.702 & 0.782 &  1 & 17 &  3  & 16.19 & 1.80 & -0.655 & 3 & sj \\
  52555 &  9.815 & 0.734 & 0.798 &  5 &  4 &  2  & 13.88 & 1.53 & -0.359 & 3 & s  \\
  56837 &  8.492 & 0.670 & 0.775 & 25 & 33 & 21  & 26.74 & 1.06 & -0.319 & 2 & j  \\
  57350 &  8.818 & 0.756 & 0.845 & 23 & 32 & 20  & 25.67 & 0.94 & -0.315 & 3 & j  \\
  58401 &  8.912 & 0.816 & 0.911 &  0 &  1 &  5  & 31.35 & 1.05 & -0.404 & 3 & s  \\
  59014 & 10.102 & 0.779 & 0.976 & 23 & 33 & 21  & 17.27 & 1.47 & -0.319 & 3 & j  \\
  60853 &  9.446 & 0.976 & 1.108 &  4 &  6 &  1  & 35.30 & 1.20 & -0.463 & 3 & sj \\
  62607 &  8.139 & 0.695 & 0.762 &  8 &  4 & 16  & 30.12 & 0.91 & -0.477 & 3 & sj \\
  64386 &  9.905 & 0.605 & 0.751 & 23 & 32 & 19  & 14.30 & 1.92 & -0.692 & 1 & j  \\
  65040 &  9.769 & 0.661 & 0.779 &  7 &  0 &  0  & 15.43 & 1.32 & -0.602 & 2 & sj \\
  67655 &  7.981 & 0.659 & 0.764 &  3 &  4 &  6  & 40.02 & 1.00 & -0.839 & 3 & s  \\
  69201 &  9.680 & 0.782 & 0.916 &  0 &  0 &  4  & 20.95 & 1.65 & -0.517 & 2 & s  \\
  72998 &  9.518 & 0.722 & 0.803 &  4 &  2 &  4  & 20.29 & 1.51 & -0.383 & 2 & s  \\
  76701 &  8.644 & 0.693 & 0.752 &  0 &  4 &  5  & 24.63 & 1.48 & -0.355 & 3 & s  \\
  78775 &  6.684 & 0.783 & 0.811 & 23 & 32 & 19  & 69.61 & 0.57 & -0.525 & 1 & j  \\
  79576 &  8.670 & 0.662 & 0.734 &  6 &  3 & 30  & 26.04 & 1.34 & -0.668 & 3 & s  \\
  81046 & 10.186 & 0.774 & 0.796 &  5 &  0 &  3  & 16.70 & 1.94 & -0.938 & 3 & sj \\
  81294 & 10.360 & 0.957 & 1.090 &  0 &  4 & 12  & 25.54 & 2.27 & -0.326 & 2 & sj \\
  83990 &  7.384 & 0.882 & 0.975 &  8 &  7 &  5  & 73.07 & 0.91 & -0.402 & 3 & s  \\
  85235 &  6.463 & 0.789 & 0.795 & 23 & 32 & 19  & 78.14 & 0.51 & -0.336 & 1 & j  \\
  91381 &  8.459 & 0.724 & 0.779 & 23 & 32 & 19  & 28.26 & 1.03 & -0.402 & 1 & j  \\ 
  91605 &  8.556 & 0.925 & 1.037 & 46 & 54 &109  & 41.88 & 1.59 & -0.317 & 1 & j  \\
  93341 & 10.140 & 0.752 & 0.788 & 25 & 35 & 29  & 15.62 & 1.83 & -0.737 & 1 & j  \\
  93623 &  9.619 & 0.674 & 0.768 &  4 &  4 &  1  & 16.45 & 1.41 & -0.527 & 4 & s  \\
  95059 & 10.566 & 0.910 & 0.991 &  3 & 11 & 16  & 18.16 & 1.88 & -0.338 & 2 & sj \\
  98792 &  7.308 & 0.820 & 0.924 & 23 & 32 & 22  & 64.17 & 0.85 & -0.325 & 1 & j  \\
  99651 &  8.671 & 0.723 & 0.838 &  1 &  1 &  1  & 27.33 & 1.21 & -0.700 & 4 & sj \\
 104289 & 10.250 & 0.692 & 0.777 &  0 &  4 &  3  & 14.64 & 1.84 & -0.601 & 4 & s  \\
 105905 &  8.664 & 0.920 & 1.018 &  1 &  3 &  1  & 43.12 & 1.17 & -0.359 & 4 & s  \\
 107314 &  9.526 & 0.789 & 0.852 & 23 & 33 & 20  & 22.39 & 1.41 & -0.457 & 1 & j  \\
 115194 &  8.943 & 0.765 & 0.869 & 23 & 32 & 19  & 31.50 & 1.18 & -0.313 & 2 & j  \\ 
\enddata
\end{deluxetable}

\clearpage
\begin{deluxetable}{crcccc}
\tablecolumns{6}
\tablewidth{0pc}
\tablecaption{Derived Distance Moduli for 47 Tuc}
\tablehead{
\colhead{}       &         & \multicolumn{4}{c}{$(m-M)_{V}$ from:} \\
Sample  &    N    & \multicolumn{2}{c}{$V/(B-V)$} &  \multicolumn{2}{c}{$V/(V-I)$}}
\startdata
1 &  41 &  13.37 & $\pm$0.03 &  13.37 & $\pm$0.03 \\
2 &  13 &  13.36 & $\pm$0.07 &  13.40 & $\pm$0.07 \\
3 &  14 &  13.37 & $\pm$0.05 &  13.33 & $\pm$0.04 \\
4 &   5 &  13.40 & $\pm$0.09 &  13.35 & $\pm$0.08 \\ 
\enddata
\end{deluxetable}

\clearpage
\begin{figure}   
\plotone{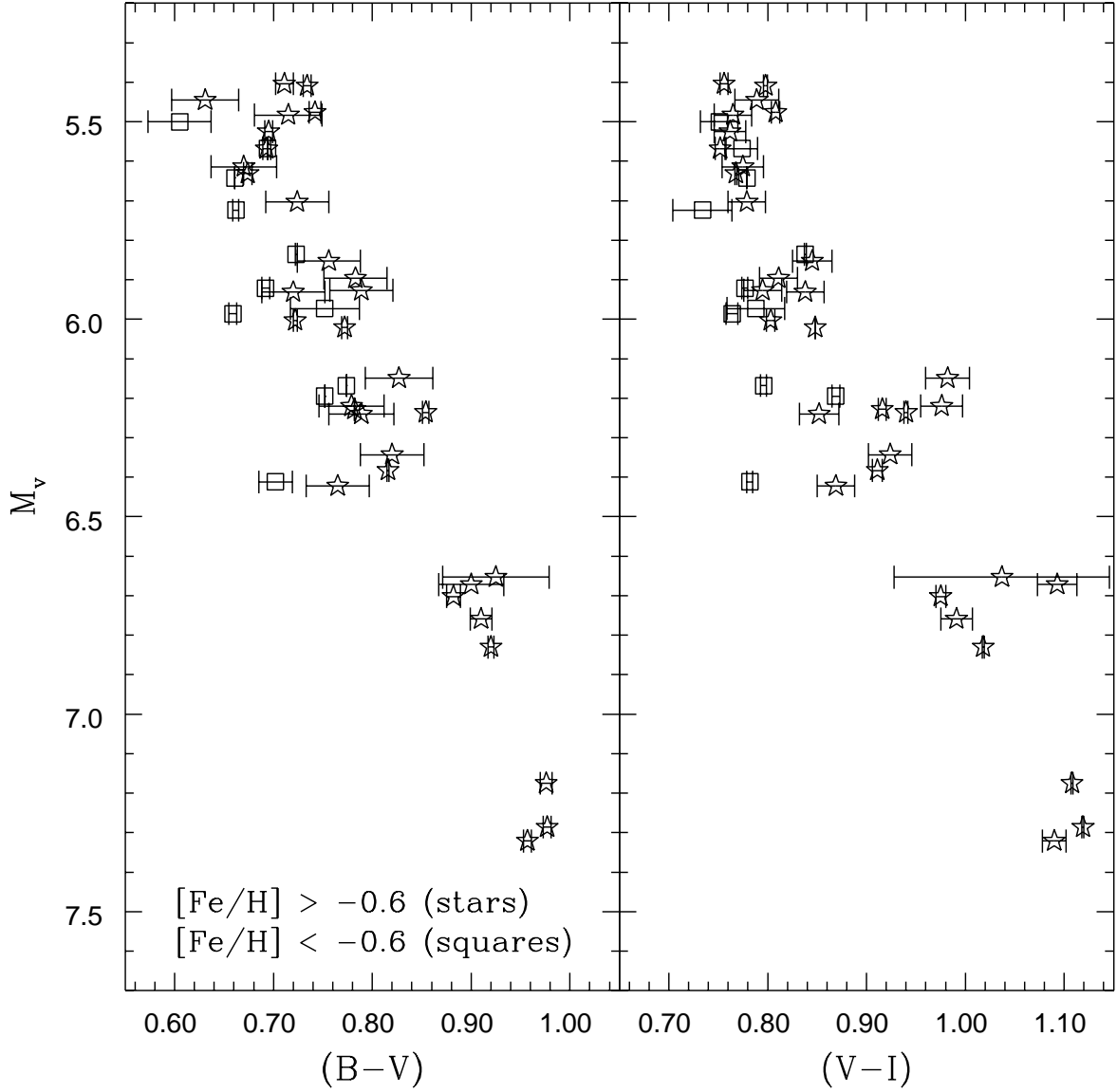}   
\caption{New photometry for 43 subdwarfs (see Table 1), divided into 2 metallicity bins:  
$[Fe/H]>-0.6$ and $[Fe/H]<-0.6$.  Error bars are photometry errors only.} 
\end{figure}   
  
\clearpage  
\begin{figure}   
\plotone{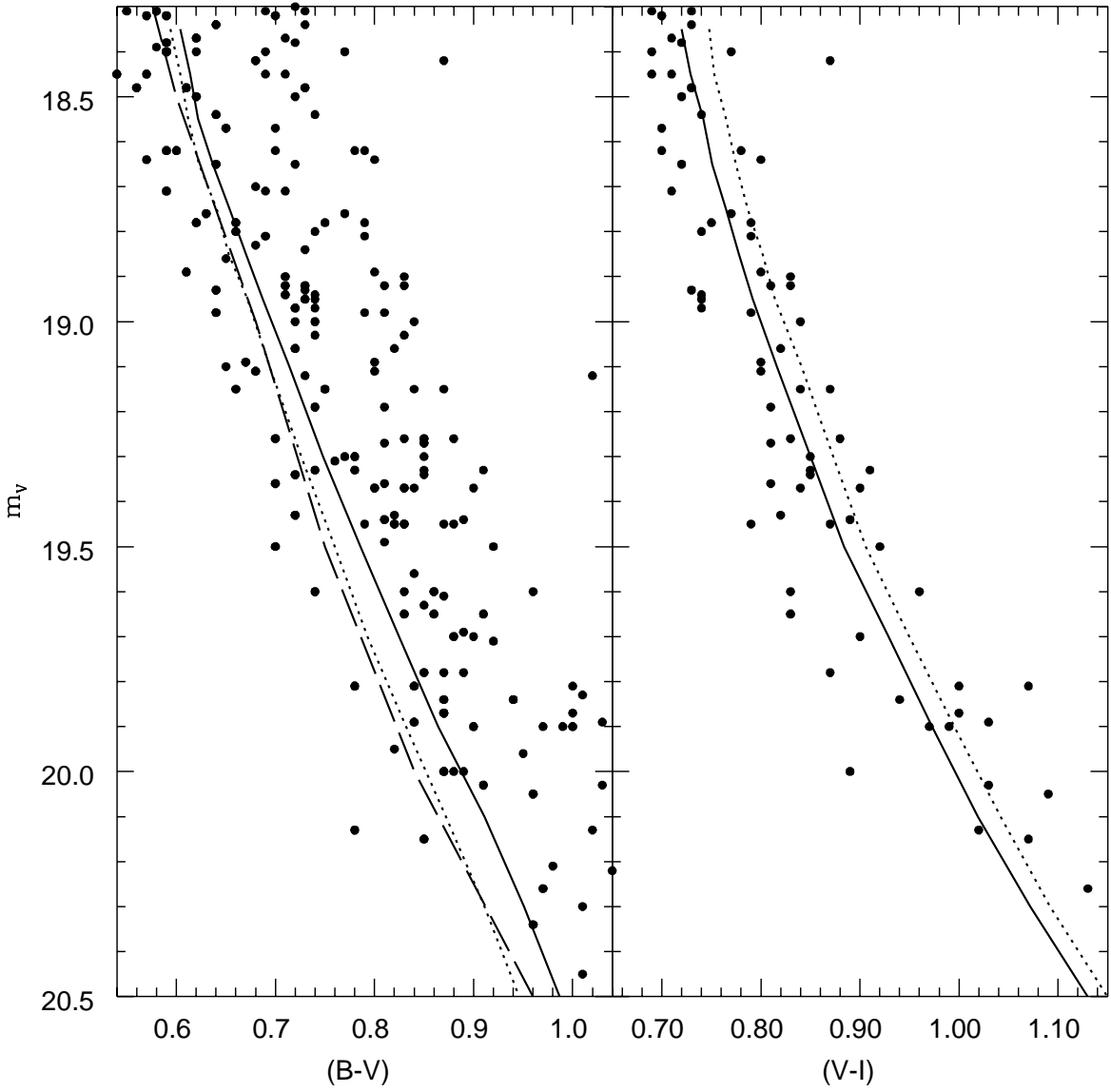}   
\caption{Comparison of 47 Tuc main-lines in $V/(B-V)$ (left panel) and $V/(V-I)$ (right panel).
Both plots show: original Kaluzny et al. (1998) main-line (dotted line), Kaluzny data recalibrated
via Stetson (2000) standards (solid line) and photoelectric data from Alcaino \& Liller (1987) (dots). 
The $V/(B-V)$ plot also shows the Hesser et al. (1987) fiducial sequence (dashed line).}     
\end{figure}    
  
\clearpage  
\begin{figure}   
\plotone{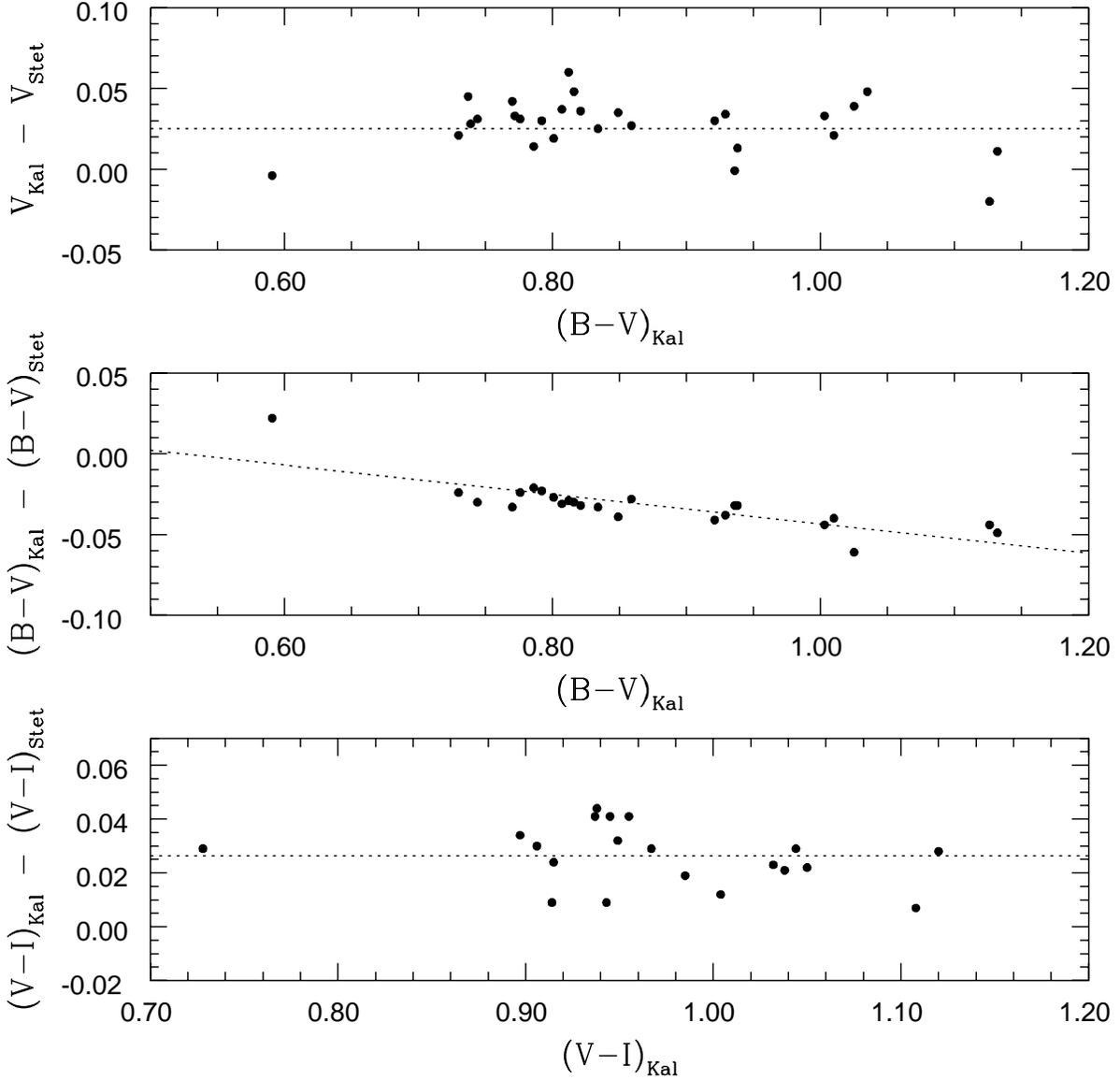}   
\caption{Comparison of Kaluzny and Stetson photometries for 47 Tuc from stars in common between
the samples.  Dotted lines show the derived offsets used to recalibrate the Kaluzny data.}    
\end{figure}   

\clearpage  
\begin{figure}   
\plotone{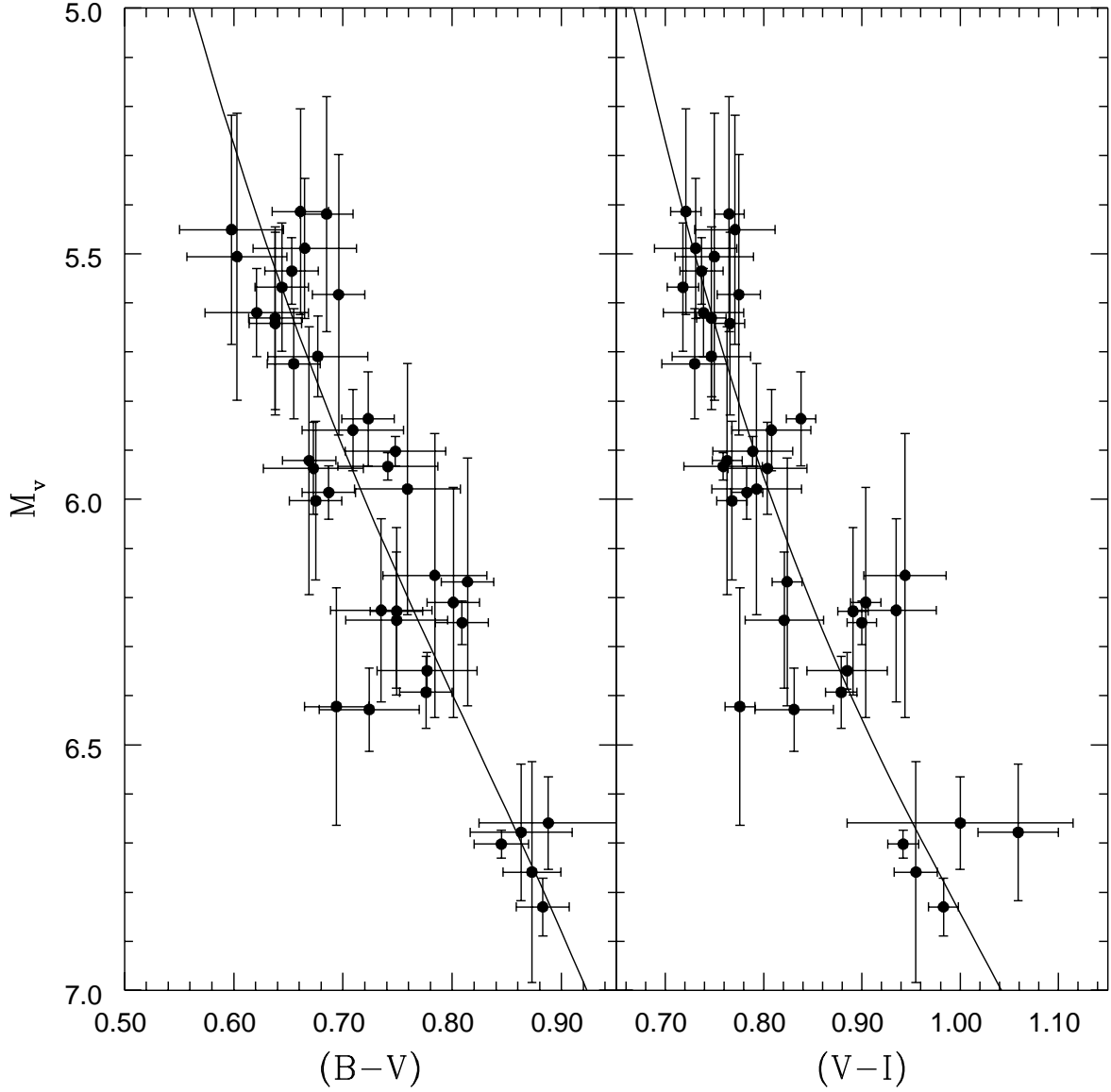}   
\caption{Whole LK-corrected subdwarf sample `shifted' to $[Fe/H]=-0.7$ and the recalibrated 47 Tuc 
main line, using our best-fit distance modulus of $(m-M)_{V}=13.37$ in both the $V/(B-V)$ and 
$V/(V-I)$ plots.}     
\end{figure}

\end{document}